\documentclass[twocolumn,aps,showpacs,preprintnumbers,amsmath,amssymb,floatfix]{revtex4}

\usepackage{graphicx}
\usepackage{booktabs} 
\usepackage{float} 

\begin{document}
\title{Doping effects of Co, Ni, and Cu in FeTe$_{0.65}$Se$_{0.35}$ single crystals}

\author{V. L. Bezusyy, D. J. Gawryluk, M. Berkowski, M. Z. Cieplak}
\affiliation{Institute of Physics, Polish Academy of Sciences,
Al. Lotnik\'{o}w 32/46, 02-668 Warsaw, Poland}

\begin{abstract}
The resistivity, magnetoresistance, and magnetic susceptibility are measured in single crystals of FeTe$_{0.65}$Se$_{0.35}$ with Cu, Ni, and Co substitutions for Fe. The crystals are grown by Bridgman's method. The resistivity measurements show that superconductivity disappears with the rate which correlates with the nominal valence of the impurity. From magnetoresistance we evaluate doping effect on the basic superconducting parameters, such as upper critical field and coherence length. We find indications that doping leads to two component superconducting behavior, possibly because of local charge depression around impurities.
\end{abstract}

\maketitle

\section{Introduction}
\label{intro}

The iron chalcogenides, FeTe$_{1-x}$Se$_x$, belong to recently discovered family of iron-based superconductors (IS), which includes also group of iron pnictides \cite{ishida,gure}. In IS superconductivity appears usually upon partial substitution of one or more elements of a magnetic parent material. In case of FeTe$_{1-x}$Se$_x$, the "end point" Fe$_{1+\delta}$Se is superconducting with superconducting transition temperature ($T_c$) of 8K, and the metallic compound Fe$_{1+\delta}$Te shows antiferromagnetic ordering but no superconductivity. Doping of Te into Se-sites increases $T_c$ to a maximum of 15 K at $x=0.5$, before decreasing it down to zero. The IS are multiband compounds. It is suggested that superconducting pairing may be mediated by spin or orbital fluctuations \cite{fluct}. The theories predict s-wave symmetry of the superconducting gap, but multiband structure allows many variations, with or without gap nodes \cite{theories}. The experiments confirm gap nodes in some of IS, but not in others.

The addition of impurities has often been used to probe the properties of superconductors. Impurities may modify the density of carriers and the band structure, may induce localized magnetic moments or influence magnetism of the host material; finally, they may enhance the scattering of carriers. In multi-band compounds the scattering may couple quasiparticle excitations on different Fermi surface sheets, with the effect on the type of superconducting order parameter which may be realized. The studies of impurity doping in various IS compounds attempt to create universal picture for all of them. For example, recent study of the critical current density in large group of iron pnictides has shown that charged impurities act as scattering centers for quasi-particles, while isovalent impurities do not \cite{kes}.

Impurity doping effects in FeTe$_{1-x}$Se$_x$ are not yet understood well. Most of the work has been done on polycrystalline specimens, for which it has been assumed that the final chemical composition is identical with the starting mixture. However, recent study of the single crystal-growth by Bridgman's method has shown that out of 17 elements, that have been examined, only three elements substituted for Fe form a single phase: Cu, Ni and Co \cite{gaw}. In the present work, we evaluate the rate of suppression of $T_c$ in the limit of small dopings of Cu, Ni, and Co into FeTe$_{0.65}$Se$_{0.35}$, and we examine other basic parameters of doped crystals. We choose FeTe$_{0.65}$Se$_{0.35}$ as a host crystal in order to obtain the best quality single-phase material. While crystals with $x=0.5$ display highest $T_c$, they show coexistence of two tetragonal phases \cite{sales}.

\section{Experimental details}
\label{sec:1}

Single crystals of nominal composition FeTe$_{0.65}$Se$_{0.35}$ and Fe$_{1-y}$M$_y$Te$_{0.65}$Se$_{0.35}$ (M = Co, Ni, Cu) are grown using Bridgman's method, from stoichiometric quantities of iron chips (3N5), tellurium powder (4N), selenium powder (pure), Co (metallic), NiSe (pure), and CuSe (4N). The growth process is described elsewhere \cite{gaw}.

The average chemical composition is checked on the natural (001) cleavage plane by field emission scanning electron microscopy (FESEM, JEOL JSM-7600F). The quantitative point analysis is done by Oxford INCA energy dispersive X-Ray spectroscopy (EDX) coupled with the SEM. X-Ray Powder Diffraction (XRPD) patterns, obtained with Siemens D5000 diffractometer, are analyzed by the Rietveld refinement method using DBWS-9807 program \cite{prog}. Major phase reflections are indexed to a tetragonal cell in the space group $P4/nmm$ (No. 129) of the PbO structural type with occupation Wyckoff's 2a site by Fe, and the 2c site by Se/Te.

The measurements of AC magnetic susceptibility are performed with magnetic field amplitude 1 Oe and frequency 10 kHz. The resistivity and magnetoresistance are taken in the $T$-range from 2 K to 300 K by standard four-probe method, using Physical Property Measurement System (Quantum Design), in magnetic fields $H$ from 0 to 14 Tesla, and directed parallel to the $ab$-plane, and to $c$-axis.

\section{Results and discussion}
\begin{figure}
\includegraphics[width=7.0cm]{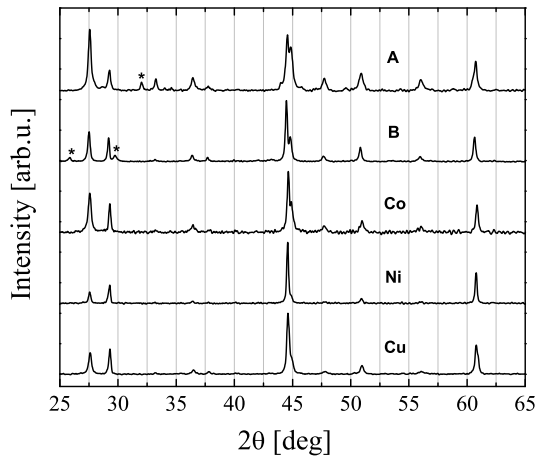}
\caption{The XRPD patterns for two crystals of FeTe$_{0.65}$Se$_{0.35}$ (A and B), and for crystals doped with nominal 1 at.\% of Co, Ni or Cu impurity. Asterisks in spectra (A) and (B) mark minority phases, most likely iron oxide inclusions.}
\label{fig:xray}
\end{figure}

Fig.\ref{fig:xray} shows XRPD spectra for two crystals with identical starting composition FeTe$_{0.65}$Se$_{0.35}$, but grown with different growth velocities, about 8 mm/h and about 1.2 mm/h (marked A and B), and three crystals with nominal 1 at.\% of impurity, Co, Ni or Cu, substituted into Fe-site, grown with the same velocity as crystal B. The full width at half maximum from $\omega$-scan on (004) diffraction line, $\Delta\omega$, equals to 1.67 arc min in crystal B, and 6 arc min in crystal A \cite{gaw}, indicating much better crystalline quality of crystal B and doped crystals.

The spectra show that all crystals are essentially single (tetragonal) phase, with small peaks from minority phases (marked by asterisks) which could be indexed to iron oxide phases. In addition, close inspection shows the presence of small inclusions of hexagonal phase of the type Fe$_7$(TeSe)$_8$. They are not visible in Fig.\ref{fig:xray}, but become apparent in transmission electron microscopy (TEM) images as described elsewhere \cite{wittlin}. Detail evaluation indicates that while average volume fraction of hexagonal inclusions is approximately the same in crystals A and B (not exceeding about 5-6\%), the size and distribution of these inclusions is different in A and B samples. While in samples A there are many small inclusions, of the size 1 to 3 nm, well separated from the major tetragonal phase, in crystal B the inclusions are fewer but larger, of the size 10 nm and more, and surrounded by the intermediate region of strained tetragonal phase.

\begin{figure}
\includegraphics[width=7.5cm]{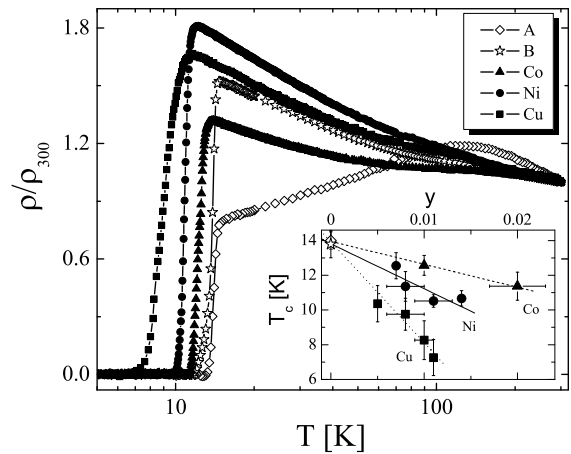}
\caption{$ab$-plane resistivity (normalized to room temperature value) for undoped crystals A and B, and for crystals doped with nominal 1 at.\% of Co, Ni or Cu impurity. The inset shows $T_c$ vs impurity content $y$. $y$ and horizontal errorbars are determined from EDX, and vertical errorbars show 10\% to 90\% resistive transition width. Dashed, solid and dotted lines are linear fits for samples doped with Co, Ni and Cu, respectively.}
\label{fig:resistivity}
\end{figure}

Different crystalline quality of crystals A and B is immediately evident during preparation of these crystals for resistivity measurement. While crystals B (and all doped crystals) are easily flaked into thin platelet-like pieces with large (several mm) platelet plane perpendicular to the $c$ axis, crystal A breaks into small irregular grains. Fig.\ref{fig:resistivity} shows $T$-dependence of the $ab$-plane resistivity $\rho$, normalized to resistivity at room temperature, ${\rho}_{300}$, for undoped (A and B) and for doped crystals with nominal 1 at\% of impurity. We see that the two undoped crystals show markedly different behaviors of $\rho(T)$. While in sample A $\rho$ decreases with decreasing $T$, indicating good metallic character, in sample B it increases with lowering of $T$, with approximate dependence $\rho \sim \log(1/T)$. Very similar low-$T$ upturn of resistivity is present also in three doped crystals. The low-$T$ upturn of resistivity is usually caused by localization of carriers. For example, similar behavior has been observed in crystals Fe$_{1+\delta}$Te$_{1-x}$Se$_x$ with $x=0.4$ \cite{liu} or $x=0.5$ \cite{rozler}, and it has been attributed to disorder-driven localization, presumably caused by the excess of Fe. However, the disorder-driven weak localization is an orbital effect which should be suppressed by the perpendicular magnetic field, causing very characteristic negative magnetoresistance effect. We have performed a preliminary measurement which suggests that this effect is absent in our samples. Therefore, the origins of the upturn must be related to some other effects. Since the crystals A and B differ by the velocity of growth, it is possible that the different $\rho (T)$ behavior is caused by differences in microstructure, such as, for example, various volume fractions of strained regions in the crystals.

Interestingly, while $\rho(T)$ is so different in crystals A and B, the $T_c$ is only slightly lower in crystal B than in crystal A. On the other hand, doping with impurities leads to substantial decrease of $T_c$. Defining as $T_c$ the temperature at which the resistance falls to half of the normal-state value, we plot in the inset to Fig.\ref{fig:resistivity} the dependence of the $T_c$ on $y$ for several crystals with small amount of impurity substituted for Fe. The vertical errorbars reflect 90\% to 10\% transition width. The $y$ values in the inset and the horizontal errorbars show average impurity content and the standard deviations, respectively, obtained from several EDX measurements performed in different points on the crystal. The straight lines fitted to the data allow to extract the rate of suppression of $T_c$ by different impurities, $dT_c/dy$. These rates are equal to about 5.8, 2.6 and 1.3 K/at.\% for Cu, Ni, and Co, respectively (see Table \ref{tab:parameters}). It is clear that they correlate with the nominal valence of the impurity, although the rate for Cu is larger than 3 times rate for Co, what may be a result of some additional factors which contribute to the $T_c$ reduction. It is possible that the main effect of the impurity may be the electron doping of the crystal, or, alternatively, that the scattering on impurities is pair-breaking, or both. Charged dopants are expected to be pair-breaking in case of so-called $s_{\pm}$ superconductivity \cite{theories}. Unfortunately, the resistivity data cannot be easily utilized to estimate the scattering rates, because (as seen in Fig.2) the upturn in resistivity does not correlate well with the $T_c$ suppression. It seems that the microstructural disorder in the crystals affects strongly resistivity while it has little effect on the $T_c (y)$ dependence.

\begin{table}
\caption{$T_c$, $H_{c2}(0)$, and $\xi(0)$ for samples A, B, and for crystals doped with Co, Ni, and Cu, $y=0.01$. $dT_c/dy$ is calculated based on data shown in the inset to Fig.\ref{fig:resistivity}.}
\label{tab:parameters}
\begin{center}
\begin{tabular}{lccccccc}
\hline\hline
\textbf{} & \textbf{A} & \textbf{B} & \textbf{Co} & \multicolumn{2}{c}{\textbf{Ni}}& \multicolumn{2}{c}{\textbf{Cu}}\\
\hline
{$T_{c} [K]$} & 14.1 & 13.9 & 12.6 & \multicolumn{2}{c}{10.7} & \multicolumn{2}{c}{8.3}\\
{$dT_{c}/dy [K/at.\%]$} & & & -1.3 & \multicolumn{2}{c}{-2.6} & \multicolumn{2}{c}{-5.8}\\
\hline
& & & & highH & lowH & highH & lowH\\
\hline
$H_{c2}^{ab}(0)$ [T] & 71 & 70 & & 39 & 22 & 22 & 5\\
$H_{c2}^c(0)$ [T] & 46 & 38 & & 23 & 14 & 17 & 3\\
\hline
$\xi_{ab}(0)$ [{\AA}] & 27 & 30 & & 38 & 48 & 44 & 105\\
$\xi_c(0)$ [{\AA}] & 17 & 16 & & 22 & 32 & 35 & 68\\
\hline
\end{tabular}
\end{center}
\end{table}

\begin{figure}
\includegraphics[width=6.5cm]{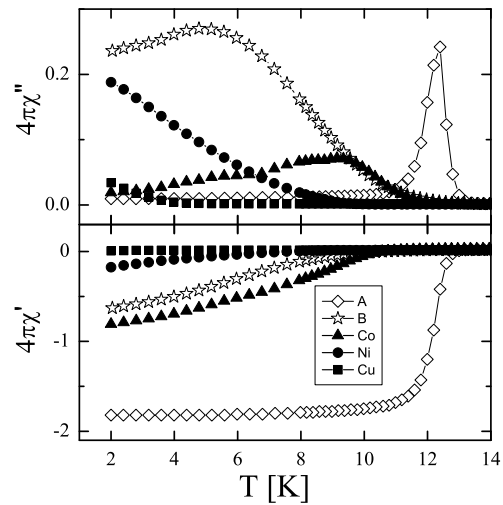}
\caption{$T$-dependence of the imaginary part (top) and the real part (bottom) of AC magnetic susceptibility measured in 1 Oe of AC field with 10 kHz in warming mode for undoped crystals A and B, and for crystals doped with nominal 1 at.\% of Co, Ni or Cu impurity (field orientation has no effect on $T_c$).}
\label{fig:susceptibility}
\end{figure}

Fig.\ref{fig:susceptibility} shows the $T$-dependence of AC susceptibility measured in field of 1 Oe, with frequency 10 kHz. The data are not corrected for demagnetizing field and therefore absolute value of real part of AC susceptibility is higher than 1. In addition, small paramagnetic background to AC susceptibility is present. All samples show diamagnetic contributions. However, while in sample A the diamagnetic contribution increases rapidly with $T$ decreasing below $T_c$, in all other samples this increase is very gradual, and the magnitude of diamagnetic signal is small. This is particularly striking for Cu-doped crystal, in which $T_c$ estimated from resistivity is quite large, 8.3 K, but in susceptibility measurements the onset of diamagnetic signal becomes apparent below 4 K only. This difference may indicate that the samples are inhomogeneous, so that the resistive transition occurs when first percolating superconducting path appears in the sample, while diamagnetic signal shows at lower $T$ when bulk superconductivity is established. Another possibility is that diamagnetic signal is smeared out by AC field due to very low magnitude of the critical current density - this may happen even at small AC field amplitude. Indeed, using magnetooptical imaging we have confirmed that the critical current density is very low in these crystals.

\begin{figure}
\includegraphics[width=7.0cm]{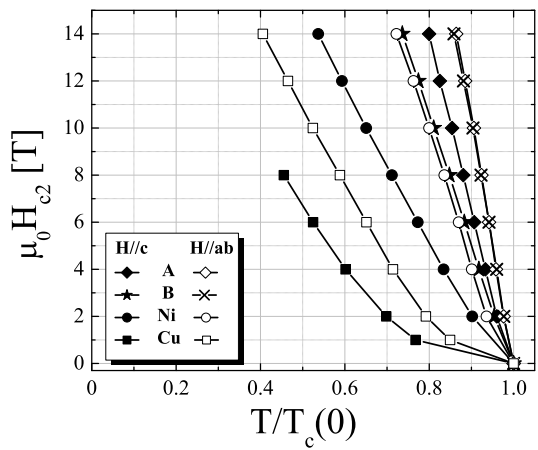}
\caption{$\mu_0 H_{c2}$ for H//ab and H//c, determined from mid-point of the resistive transition versus reduced temperature $T / T_c (H=0)$, for undoped crystals A and B, and for crystals doped with nominal 1 at.\% of Ni or Cu impurity.}
\label{fig:upper critical field}
\end{figure}

Finally, we study the suppression of the $T_c$ by external magnetic field $H$ directed parallel to the $ab$-plane, and to the $c$-axis. In Fig.\ref{fig:upper critical field} we plot upper critical field, $\mu_0 H_{c2}$, determined from the mid-point of the resistive superconducting transition, as a function of the reduced temperature, $T / T_c (H=0)$, for undoped crystals A and B, and for crystals doped with Ni and Cu ($y=0.01$). In all cases $H_{c2}^{ab}$ increases with the lowering of $T$ more steeply than $H_{c2}^c$, as have been already observed by other studies of FeSeTe \cite{gure}. Anisotropy is smaller in crystal A than in other crystals, most likely because of worse crystalline quality. Using the WHH (Werthamer-Helfand-Hohenberg) relation,  $\mu_{0}H_{c2}(0)=-0.693\mu_{0}T_{c}(dH_{c2}/dT)_{T_{c}}$, we extract the values of $H_{c2}^c (0)$ and $H_{c2}^{ab} (0)$. Note that in case of doped samples $T$-dependence of $H_{c2}$ shows an upward curvature at small $H$, particularly well pronounced in Cu-doped crystal. In this case we calculate two values of $dH_{c2}/dT$, for low and for high $H$. We than estimate Ginzburg-Landau coherence lengths using relations, $\xi_{ab}=(\Phi_0/2\pi\mu_0H_{c2}^c)^{1/2}$, and $\xi_c = \xi_{ab} H_{c2}^c / H_{c2}^{ab}$, where $\Phi_0 = 2.067\times10^{-15}$ Wb is the flux quantum. All parameters are listed in Table \ref{tab:parameters}.

Parameters for undoped samples are close to the ones which were reported \cite{gure}. Impurities reduce $H_{c2}$, and increase $\xi$. Similar trend has been observed in polycrystalline samples of FeTe$_{0.5}$Se$_{0.5}$ doped with Co \cite{shipra}. In dirty conventional superconductors scattering by impurities is expected to decrease mean free path $l$ leading to the increase of $H_{c2} \sim 1/\xi_0 l$ ($\xi_0$ is the coherence length in clean limit) \cite{gure}. This is not the case here. It is likely that the main effect of impurities is the shift of chemical potential, what masks the effect of disorder on $H_{c2}$. The studies of Hall effect and other material properties are needed to understand these results.

An interesting observation is the upward curvature in H$_{c2}$(T). A trace of this type of curvature has been recently reported in annealed crystals of Fe$_{1.01}$Te$_{0.62}$Se$_{0.38} $\cite{cao}, and attributed to multi-component response due to excess of Fe. This is similar to the behavior described for polycrystalline samples of YNi$_2$B$_2$C, in which weakly coupled grains and intergrain material are believed to contribute to two quite distinct superconducting regions \cite{rao}. It is likely that in our crystals the regions around doped impurities form areas with locally depressed charge and lowered $T_c$, quite distinct from the regions away from impurities. Such interpretation may explain the difference between $T_c$ values determined by resistivity and diamagnetism.

\section{Conclusions}
We have studied influence of Co, Ni, Cu impurities on the properties of FeTe$_{0.65}$Se$_{0.35}$ crystals grown by Brigdman's method. We find that the impurities suppress the superconducting transition temperature with different rate, which correlates with the nominal valence of the impurity. From magnetoresistance measurements we extract the upper critical fields, and coherence lengths in doped crystals. We observe some indications that doping may lead to inhomogeneous nature of superconductivity, particularly well pronounced in the crystals doped with Cu, possibly related to local depression of charge in the vicinity of impurity.

%
\section{Acknowledgments}
%
We would like to thank A. Wi\'{s}niewski, V. Domukhovski, and M. Koz{\l}owski for experimental support, and W. G. Wang for help with magnetoresistance studies. We are grateful to C. L. Chien and C. L. Broholm for their hospitality and letting us use their experimental equipment during the initial stage of this experiment at Johns Hopkins University. This work was partially supported by the EC through the FunDMS Advanced Grant of the European Research Council (FP7 Ideas), by the Polish NCS grant 2011/01/B/ST3/00462, by the NSF grant DMR 0821005 and by the European Regional Development Fund under the Operational Programme Innovative Economy NanoFun POIG.02.02.00-00-025/09.

\end{document}